\documentclass[reprint,showpacs,showkeywords,aps,prb,onecolumn,groupedaddress,longbibliography]{revtex4-1}

\usepackage[T1]{fontenc}
\usepackage{graphicx}
\usepackage{dcolumn}
\usepackage{bm}
\usepackage{hyperref}
\usepackage{color}
\usepackage{epsfig}
\usepackage{bm}

\begin{document}

\title{Parity-odd multipoles, magnetic charges and chirality in haematite ($\alpha$--Fe$_2$O$_3$) }

\author{S. W. Lovesey}
\affiliation{ ISIS Facility \& Diamond Light Source Ltd,
Oxfordshire OX11 0QX, United Kingdom}

\author{A. Rodr\'{\i}guez-Fern\'andez }
\author{J. A. Blanco }
\affiliation{Departamento de F\'{\i}sica, Universidad de Oviedo,
E--33007 Oviedo, Spain }

\date{\today}

\begin{abstract}
Collinear and canted magnetic motifs in haematite were
investigated by Kokubun $\it et$ $\it al.$ [Phys. Rev. B 78,
115112(2008)]. using x-ray Bragg diffraction magnified at the iron
K-edge, and analyses of observations led to various potentially
interesting conclusions. We demonstrate that the reported analyses
for both non-resonant and resonant magnetic diffraction at low
energies near the absorption K--edge are not appropriate. In its
place, we apply a radically different formulation, thoroughly
tried and tested, that incorporates all magnetic contributions to
resonant x-ray diffraction allowed by the established chemical and
magnetic structures. Essential to a correct formulation of
diffraction by a magnetic crystal with resonant ions at sites that
are not centres of inversion symmetry are parity-odd atomic
multipoles, time-even (polar) and time-odd (magneto-electric),
that arise from enhancement by the electric-dipole (E1) -
electric-quadrupole (E2) event. Analyses of azimuthal-angle scans
on two space-group forbidden reflections, hexagonal $(0,0,3)_h$
and $(0,0,9)_h$, collected by Kokubun $ \it et \,al.$ [Phys. Rev.
B 78, 115112(2008)] above and below the Morin temperature $(T_M =
250\, K)$, allow us to obtain good estimates of contributing polar
and magneto-electric multipoles, including the iron anapole. We
show, beyond reasonable doubt, that available data are
inconsistent with parity-even events only (E1-E1 and E2-E2). For
future experiments, we show that chiral states of haematite couple
to circular polarization and differentiate E1-E2 and E2-E2 events,
while the collinear motif supports magnetic charges.
\end{abstract}


\maketitle

\section{Introduction}

Enigmas about ichor-like haematite ($\alpha$--Fe$_2$O$_3$) and
famed lodestone, both real and concocted, have been worried and
written about from the time of Greek texts in 315 B.C., to William
Gilbert of Colchester, the father of magnetism, in the 16th. C.,
to Dzyaloshinsky in 1958 who gave a phenomenological theory of
weak ferromagnetism. Haematite is the iron sesquioxide that
crystallizes into the corundum structure (centro-symmetric
space-group $\sharp 167$, $R\bar{3}c$) in which ferric ($Fe^{3+}$,
$3d^5$) ions occupy sites 4(c) on the trigonal c-axis that are not
centres of inversion symmetry. For an extensive review of the
history and properties of haematite see, for example, Morrish
\cite{Morrish2} and Catti $\it et \, al$.\cite{Catti3}

\begin{figure}[h]
\begin{center}
\includegraphics[width=10cm,angle=90]{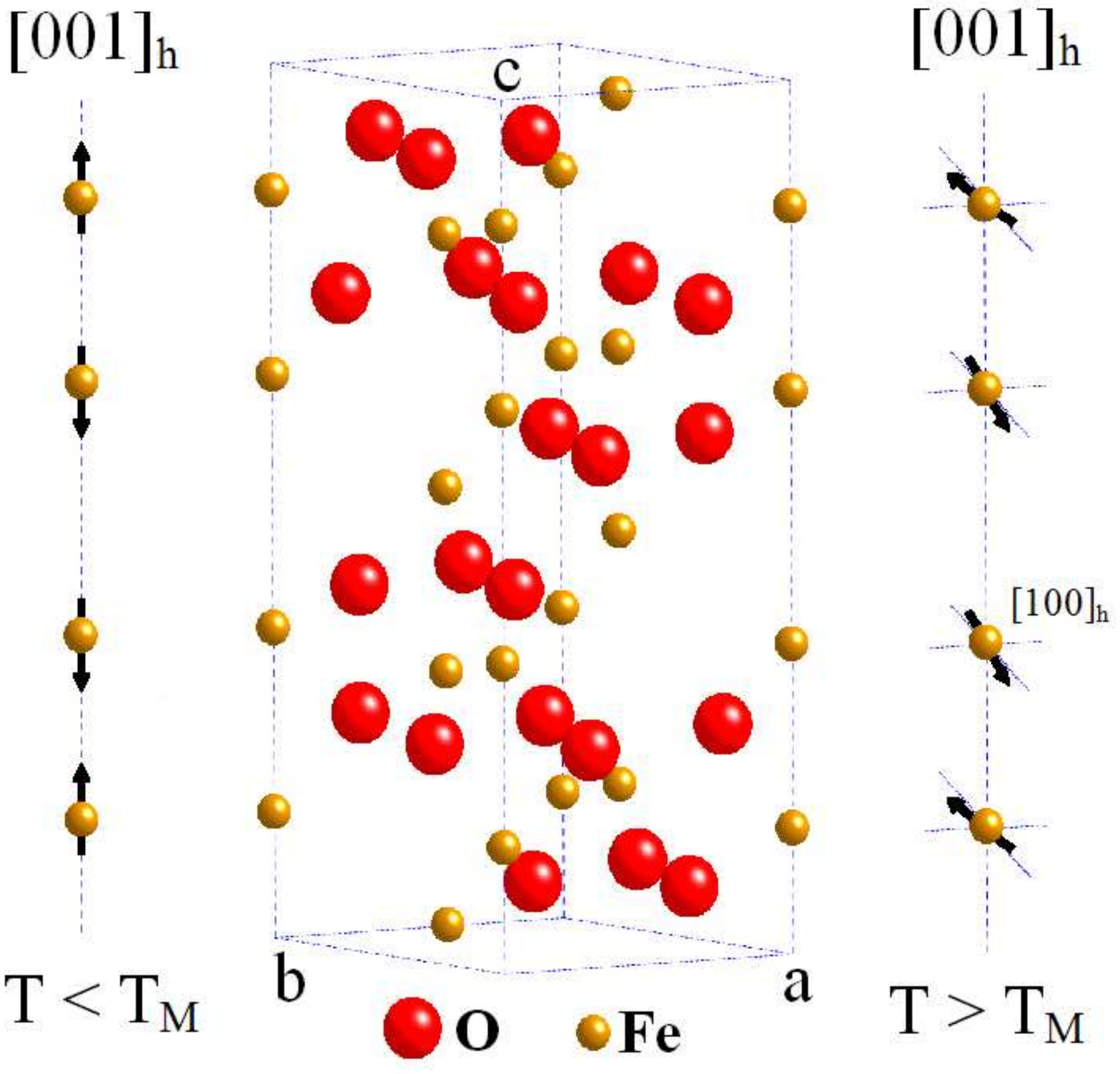}
\caption{\label{fig:figure1}Magnetic and chemical structure of
haematite, space group $R\bar{3}c$. The red and the yellow dots
represent oxygen and iron sites, respectively. The left line
denotes the magnetic motif along the $c$-axis below the Morin
temperature (Phase I). The right line denotes the motif above the
Morin temperature, where iron moments are contained in the $a$-$b$
plane (Phase II).}
\end{center}
\end{figure}

At room temperature, the motif of magnetic moments is canted
antiferromagnetism with moments in a (basal) plane normal to the
$c$-axis. Weak ferromagnetism parallel to a diad axis of rotation
symmetry, normal to a mirror plane of symmetry that contains the
$c$-axis, is created by a Dzyaloshinsky-Moriya antisymetric
interaction $  {\bf D}\cdot({\bf S_1} \times {\bf S_2})$ between
spins $ {\bf S_1} $ and $  {\bf S_2} $ and the vector ${\bf D}$ is
parallel to the $c$-axis, Dzyaloshinsky,\cite{Dzyaloshinsky4}
Moriya.\cite{Moriya5} The Morin temperature $ \rm 250$ $K$, at
which moments rotate out of the the basal plane to the $c$-axis,
may be determined from the temperature dependence of magnetic
Bragg peaks observed by neutron diffraction. Rotation of the
moments takes place in a range of $10$ $K$ in pure crystals but
the interval can be much larger, $\approx 150$ $K$, in mixed
materials.\cite{Kren6} Ultimately, moments align with the $c$-axis
and create a fully compensating, collinear antiferromagnet with an
iron magnetic moment $= 4.9$ $\mu_B$ at $77$ $K$. We follow
Dzyaloshinsky \cite{Dzyaloshinsky4} and label collinear
(low-temperature phase) and canted (room-temperature phase)
antiferromagnetism as phases I and II, respectively, see figure
\ref{fig:figure1}. In phase I haematite is not magneto-electric
unlike eskolaite $( Cr_2O_3)$, which also possesses the corundum
structure and collinear antiferromagnetism.

Finkelstein $ \it et \, al. $ \cite{Finkelstein7} and Kokubun $\it
et \, al.$ \cite{Kokubun1} studied haematite by x-ray Bragg
diffraction, with Bragg intensities enhanced by tuning the energy
of the primary x-rays to the iron K-absorption edge. In these
experiments, attention is given to Bragg reflections that are
forbidden by extinction rules for the space-group. Often called
Templeton and Templeton reflections,\cite{Templeton8} the
reflections in question are relatively weak and arise from angular
anisotropy of valence states that accept the photo-ejected
electron. Following rotation of the crystal about a Bragg
wave-vector aligned with the $c$-axis, Finkelstein $et\, al.$
\cite{Finkelstein7} observed a near six-fold periodicity of the
intensity that is traced to a triad axis of rotation symmetry that
passes through sites occupied by resonant, ferric ions. In general
by measuring intensities, collected at space-forbidden
reflections, we can obtain information of high-order multipoles
existing in the materials such as magnetic charge (or magnetic
monopole),\cite{Lovesey9} electric dipole,\cite{Fernandez10}
anapole,\cite{Lovesey11, Fernandez12} quadropole,\cite{Wilkins13}
octupole\cite{Paixao14, Lovesey15} and
hexadecapole.\cite{Tanaka16, Fernandez17} Therefore, these weak
reflections are extremely sensitive to charge, orbital and spin
electron degrees of freedom and haematite is no
exception.\cite{Lovesey18}

We apply an atomic theory of resonant Bragg diffraction formulated
for the corundum structure,\cite{Lovesey19} to data gathered by
Kokubun $\it et \, al.$ \cite{Kokubun1} at forbidden reflections
$( 0,0,l)_h$ with $l = 3(2n + 1)$ and infer from available data
relative values of atomic multipoles of the resonant ion. A
successful story emerges with scattering represented by a mixture
of parity-even and parity-odd (even or odd with respect to the
inversion of space) multipoles at sites in the structure occupied
by resonant iron ions, which are not centres of inversion
symmetry. Parity-odd multipoles arise in a resonant event using
the electric dipole (E1) and electric quadrupole (E2) -
corresponding multipoles are labelled polar (time-even) or
magneto-electric (time-odd) - while parity-even multipoles arise
from E1-E1 and E2-E2 events. A chiral state of haematite is
demonstrated by a predicted coupling of resonant intensity to
circular polarization (helicity) in the primary beam, and the
effect also differentiates between E1-E2 and E2-E2 events. The two
parity-odd multipoles of rank zero correspond to chirality and
magnetic charge \cite{Lovesey20, Lovesey21} and both pseudo-scalar
monopoles are present in the electric dipole-magnetic dipole
(E1-M1) amplitude for resonant scattering by haematite in phase I.

Our communication is arranged as follows. Section 2 contains
essential information and definitions. Unit-cell structure factors
for Bragg diffraction enhanced by E1-E1, E1-E2 and E2-E2 listed in
an Appendix are exploited in Sections 3 and 4, which report the
successful analysis of Bragg diffraction data gathered on
haematite at room temperature and $150$ $K$, well below the Morin
transition.  Thereafter, in Section 5, there are simulations of
resonant intensity induced by circular polarization in the primary
x-ray beam which signals existence of a chiral state. Section 6
addresses magnetic charge found in the E1-M1 structure factor, and
not visible in a dichroic signal. A discussion of findings in
Section 7 concludes the communication.

\section {Basics}

There are four contributions to the amplitude of photons scattered
by electrons calculated in the first level of approximation in the
small quantity $(E/mc^2)$, where E is the energy of the primary
photon, namely, Thomson scattering, spin scattering and two
contributions with virtual intermediate states, one of which may
become large when E coincides with an atomic resonance. Of
particular interest with magnetic samples is a celebrated
reduction of the amplitude, derived by de Bergevin and
Brunel,\cite{Bergevin22} which occurs at large E. In this limit,
all three contributions excluding Thomson scattering add to give
so-called magnetic, non-resonant scattering made up simply of spin
and orbital magnetic moments. De Bergevin and Brunel's result is
not valid at low energies, and certainly not below an atomic
resonance, as is at once obvious from steps in its
derivation.\cite{Lovesey23}

In an analysis of x-ray Bragg diffraction data for haematite
collected at space-group forbidden reflections we use the spin and
resonant contributions to the scattering amplitude. The spin
contribution $G^s = i(E/mc^2)({\bf e} \times {\bf e'}) \cdot
F_s(k)$ with ${\bf k} = {\bf q} - {\bf q'}$ where ${\bf e}$ and
${\bf q}$ (${\bf e'}$ and ${\bf q'}$) are, respectively, the
polarization vector and wave-vector of the primary (secondary)
photon, and the Bragg angle $\theta$ that appears in structure
factors for resonant scattering is defined by ${\bf q} \cdot {\bf
q'} = q^2 cos(2\theta).\,F_s(k)$ is the unit-cell structure factor
for spin magnetic moments. The measured energy profiles of
reflections $(0,0,3)_h$ and $(0,0,9)_h$ show a single resonance in
the pre-edge region, devoid of secondary structure, which is
modelled by a single oscillator centred at an energy  $\Delta =
7.105$ $keV$ with a width $\Gamma$, to an excellent
approximation.\cite{Kokubun1} In this instance, the resonant
contribution to scattering is represented by $d(E) F_{\mu'\nu}$
where $d(E) = \Delta/[E -\Delta + i\Gamma ]$ and $F_{\mu'\nu} $ is
a unit-cell structure factor for states of polarization
$\mu'$(secondary) and $\nu$ (primary). We follow the standard
convention for orthogonal polarization labels $\sigma$ and $\pi$;
$\sigma$ normal to the plane of scattering and, consequently,
$\pi$ in the plane. Unit-cell structure factors listed in an
Appendix are derived following steps for the corundum structure
found in Lovesey $\it et \, al.$\cite{Lovesey19} The generic form
of our Bragg scattering amplitude for haematite at a space-group
forbidden reflection (no Thomson scattering) is,

\begin{equation}
 G_{\mu'\nu}  (E) = G^s_{\mu'\nu}+ \rho \, d(E) \, F_{\mu'\nu},
\label{Eq1}
\end{equation}

where $\rho$ is a collection of factors, which include radial
integrals for particular resonance events, that are provided in an
Appendix.

Atomic multipoles  $\langle T^K_Q \rangle$  in parity-even
structure factors, for E1-E1 and E2-E2 events, have the property
that even rank K are time-even (charge) and odd rank K are
time-odd (magnetic). For enhancement at the K-absorption edge, all
parity-even atomic multipoles relate to orbital degrees of freedom
in the valence shell - spin degrees of freedom are absent Lovesey.
\cite{Lovesey24} Thus, for enhancement at the K-absorption edge,
multipoles $\langle T^K_Q \rangle$ with odd K are zero if the
ferric, $3d^5$ (electron configuration $ ^6$S) of the iron ion is
fully preserved in haematite. The measured iron magnetic moment
$4.9$ $\mu_B$ at $77$ $K$ indicates that the orbital magnetic
moment is small and likely no more than  $ \approx 2 \% $ of the
measured moment.\cite{Catti3, Kren6}

\section {Phase I}

We report first our analyses of data gathered by Kokubun $\it et
\, al.$\cite{Kokubun1} on haematite at $150$ $K$. With $100\%$
incident $\sigma$-polarization and no analysis of polarization in
the secondary beam, the measured intensity of a Bragg reflection
is proportional to,

\begin{equation}
 I =  \mid G_{\sigma'\sigma}(E)\mid ^2 +  \mid G_{\pi'\sigma}(E)\mid^2,
\label{Eq2}
\end{equation}

For a collinear antiferromagnet, in expression (\ref{Eq1}) for
G$_{\mu'\nu}$(E) one has $G^s_{\sigma'\sigma}  = 0$ and in the
channel with rotated polarization,

\begin{equation}
 G^s_{\pi'\sigma}  = 4 \, sin(\theta ) \, sin(\varphi l) \, (E/mc^2) \, f_s(k) \, \langle S^z\rangle ,
\label{Eq3}
\end{equation}

where  $\varphi = -37.91$  deg., the Bragg angle  $\theta = 10.96$
deg. $(34.77$ $deg.)$ for a Miller index $l = 3$ $(9)$, $\langle
S^z\rangle \leq 5/2$ is the spin moment and $f_s(k)$ is the spin
form factor with $f_s(0) = 1$. Note that  $\mid
G^s_{\pi'\sigma}\mid^2 \propto sin^2(\theta)$ above is not the
expression in equation (20) in Ref. (\onlinecite{Kokubun1}), which
is derived by use of an abridged scattering amplitude that is not
valid in the experiment.\cite{Bergevin22}

At resonance, the spin contribution $G^s_{\pi'\sigma}$  is
suppressed compared to the resonant contribution by a factor
$\Gamma / \Delta \approx 10^{-4}$ and it may safely be neglected.

Confrontations between our theoretical expressions for the
azimuthal-angle dependence of Bragg intensity with corresponding
experimental data reported in Ref. (\onlinecite{Kokubun1}) reveal
a $30$ deg. mismatch of origins in the azimuthal angle. Our origin
$\psi = 0$ has the $a$-axis normal to the plane of scattering
\cite{Lovesey19} whereas Kokubun $\it et \, al.$\cite{Kokubun1}
specify an origin such that the $a$-axis is parallel to ${\bf q} +
{\bf q'}$ giving a nominal mismatch in the origin of $\psi$ ,
between theory and experiment, of $90$ deg. The actual mismatch,
$30$ deg., revealed by our analysis of data is likely to arise in
the experiments by mistakenly using for reference a basal plane
Bragg reflection off-set by $60$ deg. In this and the following
section we reproduce data as a function of $\psi$ off-set by $30$
deg. compared to data reported in Figures $5$ and $10$ in Ref.
(\onlinecite{Kokubun1}).

\begin{figure}[h]
\begin{center}
\includegraphics[width=10cm]{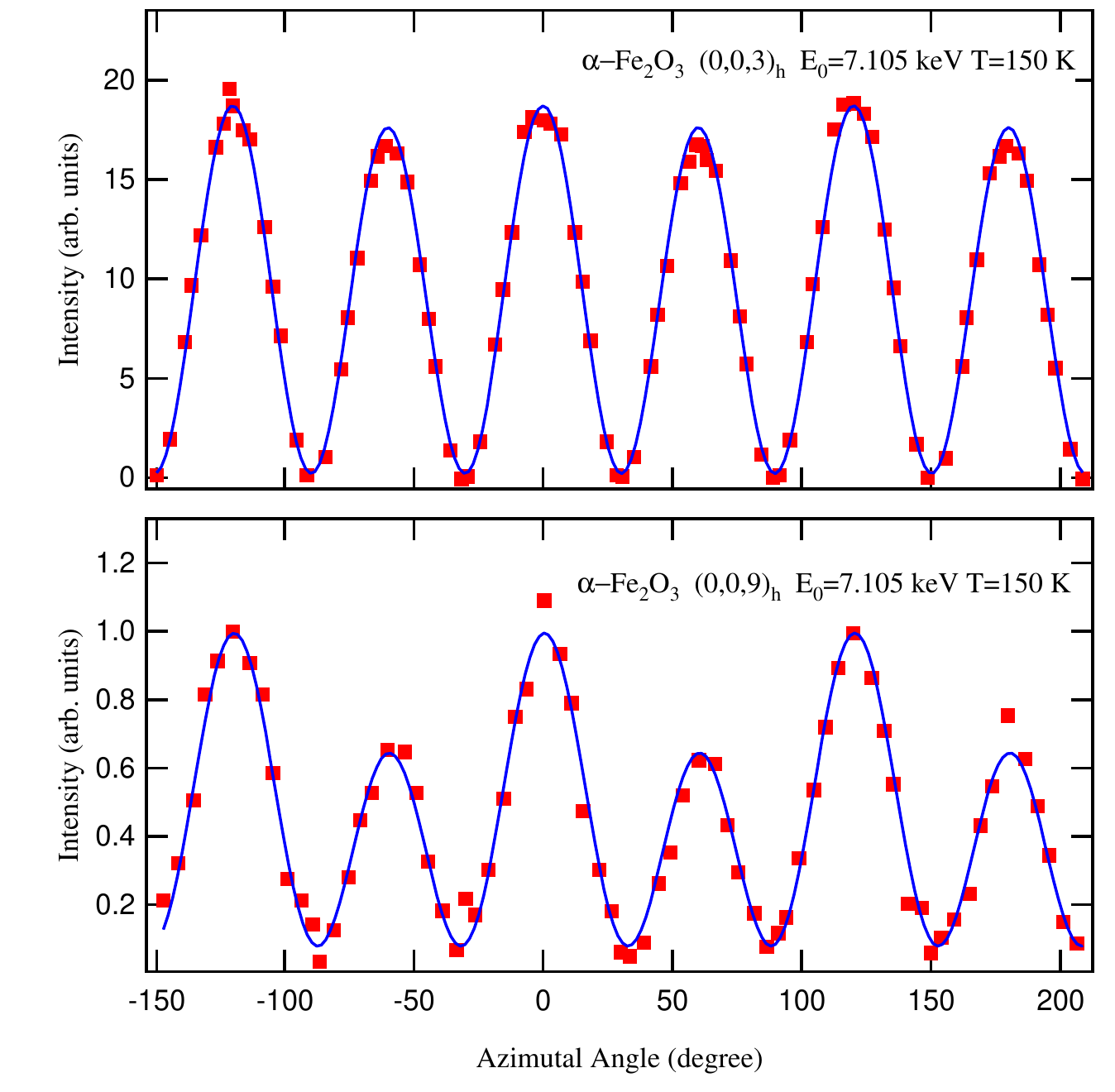}
\caption{\label{fig:figure2} Azimuthal-angle dependence of
intensity of Bragg reflections $(0,0,l)_h$ with $l = 3$ and $l =
9$ for phase I $(150 K)$. Continuous curves are fits to structure
factors for E1-E2 and E2-E2 events with magnetic (time-odd)
parity-even multipoles set to zero. Inferred relative atomic
multipoles are listed in Table \ref{tab:table1}. Experimental data
is taken from Kokubun $\it et \, al.$\cite{Kokubun1}}
\end{center}
\end{figure}

In light of the established negligible orbital magnetism in
haematite, parity-even, time-odd atomic multipoles $(K =\, 1\, \&
\, 3)$ are set equal to zero. Looking in the Appendix one finds
$F_{\mu'\nu}(E1-E1) = 0$. Additionally, $F_{\sigma'\sigma}(E2-E2)
= 0$ and $F_{\pi'\sigma}(E2-E2)$ produces Templeton - Templeton
scattering proportional to $[ \langle
T^4_{+3}\rangle'cos(3\psi)]$, where $\psi$ is the azimuthal angle.
Inspection of data for phase I reproduced in figure
\ref{fig:figure2} shows that an E2-E2 event on its own is not an
adequate representation. The missing modulation is produced by the
E1-E2 event that introduces a polar quadrupole $\langle
U^2_0\rangle$ in phase with the parity-even
hexadecapole.\cite{Dmitrienko25} Figure \ref{fig:figure2} displays
satisfactory fits of $\{ \mid F_{\sigma'\sigma}\mid ^2 + \mid
F_{\pi'\sigma} \mid ^2\} $, using equal measures of E1-E2 and
E2-E2 events, to data from azimuthal-angle scans performed at
reflections $(0,0,l)_h $ with $l = 3$ and $9$. The influence of
the polar quadrupole is very notable for $l = 9$ because for this
Miller index the hexadecapole is suppressed, with the ratio at $l
= 9$ to $l = 3$ of $tan(\varphi l)$ equal to $0.15$. Relative
values of multipoles inferred from fits to the low temperature
data are gathered in table \ref{tab:table1}. Values of $\langle
T^4_{+3}\rangle'$ and $\langle U^2_0\rangle$ in phase I are found
to be of one sign and in the ratio 20 : 1, with near equal
magnitudes of the polar quadrupole and magneto-electric octupole,
$\langle G^3_{+3}\rangle '$. If $ \mid \rho(E2-E2)/
\rho(E1-E2)\mid \approx 1.0$, as suggested by our estimate,
magneto-electric multipoles are $\approx 5 \%$ of the dominant
parity-even hexadecapole, $\langle T^4_{+3}\rangle'$.

Without polarization analysis, it does not seem possible from
azimuthal-angle scans to distinguish between E1-E2 and E2-E2
events.  However, as shown in Section 5, the two events can be
distinguished with circularly polarized x-rays.

\begin{figure}[h]
\begin{center}
\includegraphics[width=10cm]{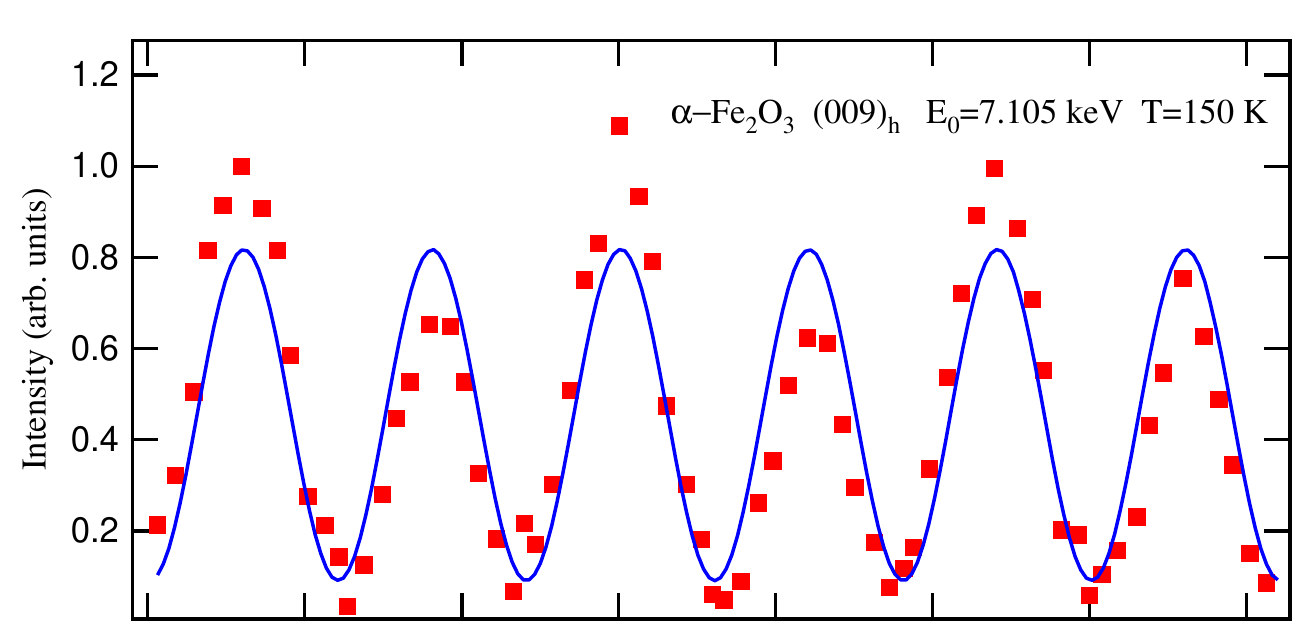}
\\
\includegraphics[width=10cm]{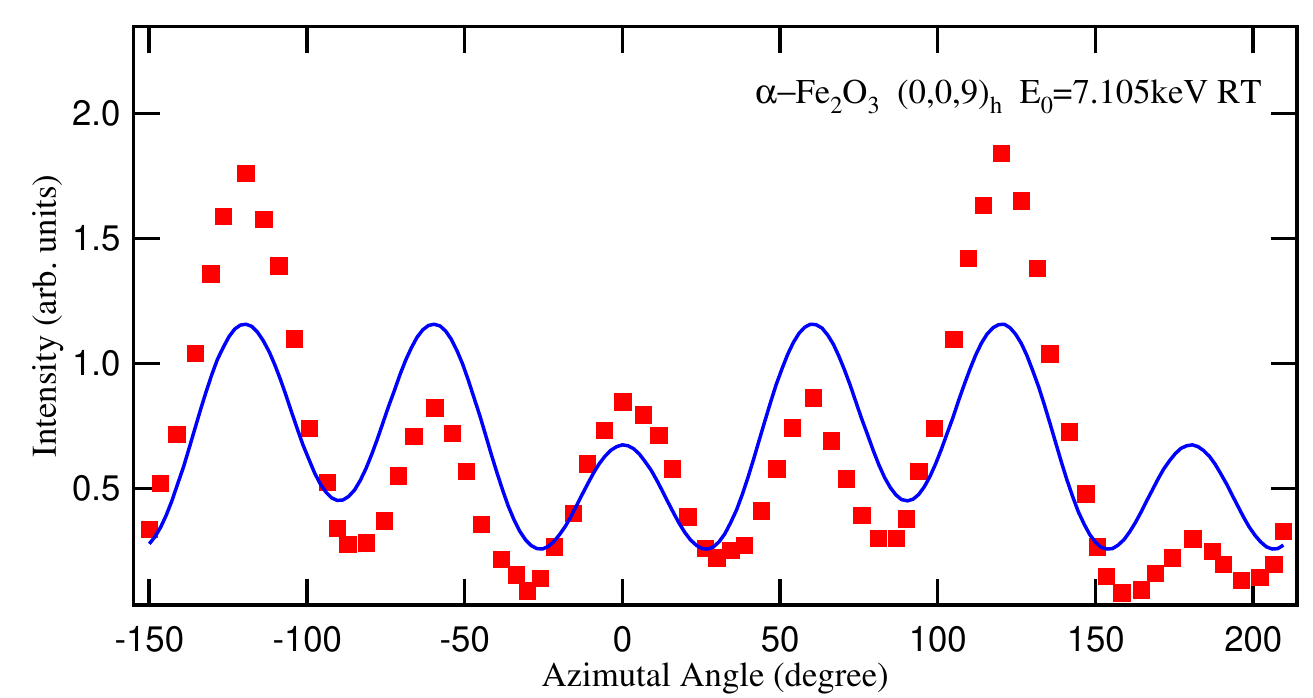}
\caption{\label{fig:figure3} Azimuthal-angle dependence of
intensity of the Bragg reflection $(0,0,9)_h$ for phases I $(150
K)$ and II (room temperature). Continuous curves are fits to
parity-even structure factors E1-E1 and E2-E2 including all
magnetic multipoles. Experimental data taken from Kokubun $\it et
\, al.$\cite{Kokubun1} appears also in figure \ref{fig:figure2}
and \ref{fig:figure4}.}
\end{center}
\end{figure}

The failure of pure parity-even structure factors E1-E1 plus E2-E2
to explain the data is most pronounced for $l = 9$. To illustrate
the extent of the failure, figure \ref{fig:figure3} displays a fit
to intensity at $l = 9$ with an amplitude made of equal amounts of
E1-E1 and E2-E2 unit-cell structure factors, and the quality of
the fit is clearly inferior to the one shown in figure
\ref{fig:figure2}.

\begin{table}[h]
\caption{\label{tab:table1}Relative values of atomic multipoles
for collinear antiferromagnetism in phase I (at  $\approx$ 100\,K
below the Morin transition) and canted antiferromagnetism in phase
II (room temperature). The magnitude of the dominant hexadecapole,
$\langle T^4_{+3}\rangle'$, is set to $+$ 10.00. The estimate
$\langle U^2_0\rangle = + 0.50$ inferred by fits to data for phase
I is also used in analysis of data for phase II. Values for other
multipoles are inferred by fitting to data equal measures of E1-E2
and E2-E2 structure factors listed in an Appendix, with time-odd
figures (magnetic) multipoles in E2-E2 set to zero. Fits are
displayed in figures \ref{fig:figure2} and \ref{fig:figure4}.
\\
With our definition, real $\langle ...\rangle$' and imaginary
$\langle ...\rangle$'' parts of a multipole are defined through
$\langle G^K_Q\rangle = \langle G^K_Q\rangle' + i \langle
G^K_Q\rangle ''$ with $\langle G^K_Q \rangle^* = (-1)^Q \langle
G^K_{-Q} \rangle$, and identical relations for the other two
multipoles, $\langle T^K_Q\rangle$ and $\langle U^K_Q\rangle $.
All multipoles with projection Q = 0 are purely real. Using radial
integrals from an atomic code factors in equation (\ref{Eq1}) are
in the ratio $\rho(E2-E2)/ \rho(E1-E2) \approx - 0.98 $, which is
no more than a guide to the actual value in haematite. This ratio
is not eliminated in listed values of multipoles.}
\begin{ruledtabular}
\begin{tabular}{@{}lcc}

    Multipole                       &     Phase I    &   Phase II      \\
\colrule
 $\langle G^1_{+1} \rangle '  $     &     $ -     $  &  $0.50 (2)$     \\
 $\langle G^2_0 \rangle       $     &   $0.11 (2) $  &     $ - $       \\
 $\langle G^2_{+1} \rangle '' $     &     $ -     $  & $-0.38 (3)$     \\
 $\langle G^3_{+1} \rangle '  $     &     $ -     $  & $ 1.07 (6) $    \\
 $\langle G^3_{+3} \rangle '  $     &   $0.41 (2) $  &  $2.45 (5)$     \\

\end{tabular}
\end{ruledtabular}
\end{table}

\section {Phase II}
In this phase, above the Morin transition, iron magnetic moments
lie in a plane normal to the $c$-axis. We choose orthonormal
principal-axes ($x$, $y$, $z$) with the $x$- and $z$-axes parallel
to the crystal $a$- and $c$-axes, respectively. The crystal
$a$-axis is parallel to a diad axis of rotation symmetry, normal
to the mirror plane that contains the trigonal $c$-axis.

The spin contribution $G^s_{\sigma ' \sigma} = 0$, while the
corresponding $\pi'\sigma$-scattering amplitude can be different
from zero and, notably, it depends on azimuthal angle. We find,

\begin{equation}
 G^s_{\pi'\sigma} = 4 \, cos(\psi) \, cos(\theta) \, sin(\varphi l) \,(E/mc^2) \,f_s(k) \, \langle S^y \rangle,
\label{Eq4}
\end{equation}

and  $ \mid G^s_{\pi ' \sigma} \mid ^2 \propto cos^2(\theta)$ from
eq. (\ref{Eq4}) is not the same as the corresponding result,
equation (19) in Ref. (\onlinecite{Kokubun1}) for reasons spelled
out in Section 3.

Away from a resonance, the result (\ref{Eq4}) predicts a two-fold
periodicity of intensity as a function of azimuthal angle, which
is in accord with observations in Ref. (\onlinecite{Kokubun1}).
Spin moment in the mirror plane  $\langle S^y\rangle$  is close to
$5/2$ while spontaneous magnetization, directed along a diad axis,
is   $ \approx 0.02 \%$ of the nominal value. From (\ref{Eq3}) and
(\ref{Eq4}) we see that the ratio of  $\mid G^s_{\pi ' \sigma}
\mid ^2$ for phases I and II depends on $tan^2(\theta )$ which
takes the value $0.04 (0.48)$ for $l = 3 (l = 9)$. For $l = 3$,
Kokubun $\it et \, al.$ \cite{Kokubun1} report intensity between
$150$ K (phase I) and $300$ K (phase II). Starting from $\approx
210$ K a large increase of intensity is observed over an interval
of $\approx 40$ K. Rotation of magnetic moments from the $c$-axis
to basal plane, between phases I and II, takes place in a range of
$10$ K in pure crystals but the interval can be larger in mixed
materials as commented above.

Slightly away from the resonance, interference between the
non-resonant, spin contribution (\ref{Eq4}) and $d(E)F_{\pi '
\sigma} $ may enhance intensity in a Bragg peak if  $(E - \Delta )
[G^s_{\pi ' \sigma} / (F_{\pi ' \sigma})'] \rangle 0$. We find
$[G^s_{\pi ' \sigma}/(F_{\pi ' \sigma})']$ is of one sign for $l =
3$ and $l = 9$ provided that $f_s(k)$, the spin form factor, is of
one sign. At face value this finding is not at one with Kokubun
$\it et \, al.$\cite{Kokubun1} who discuss a sighting of slight
enhancement of the intensity on the low-energy side of the
resonance for $l = 9$ that is apparently absent, or completely
negligible, for $l = 3$.

\begin{figure}[h]
\begin{center}
\includegraphics[width=10cm]{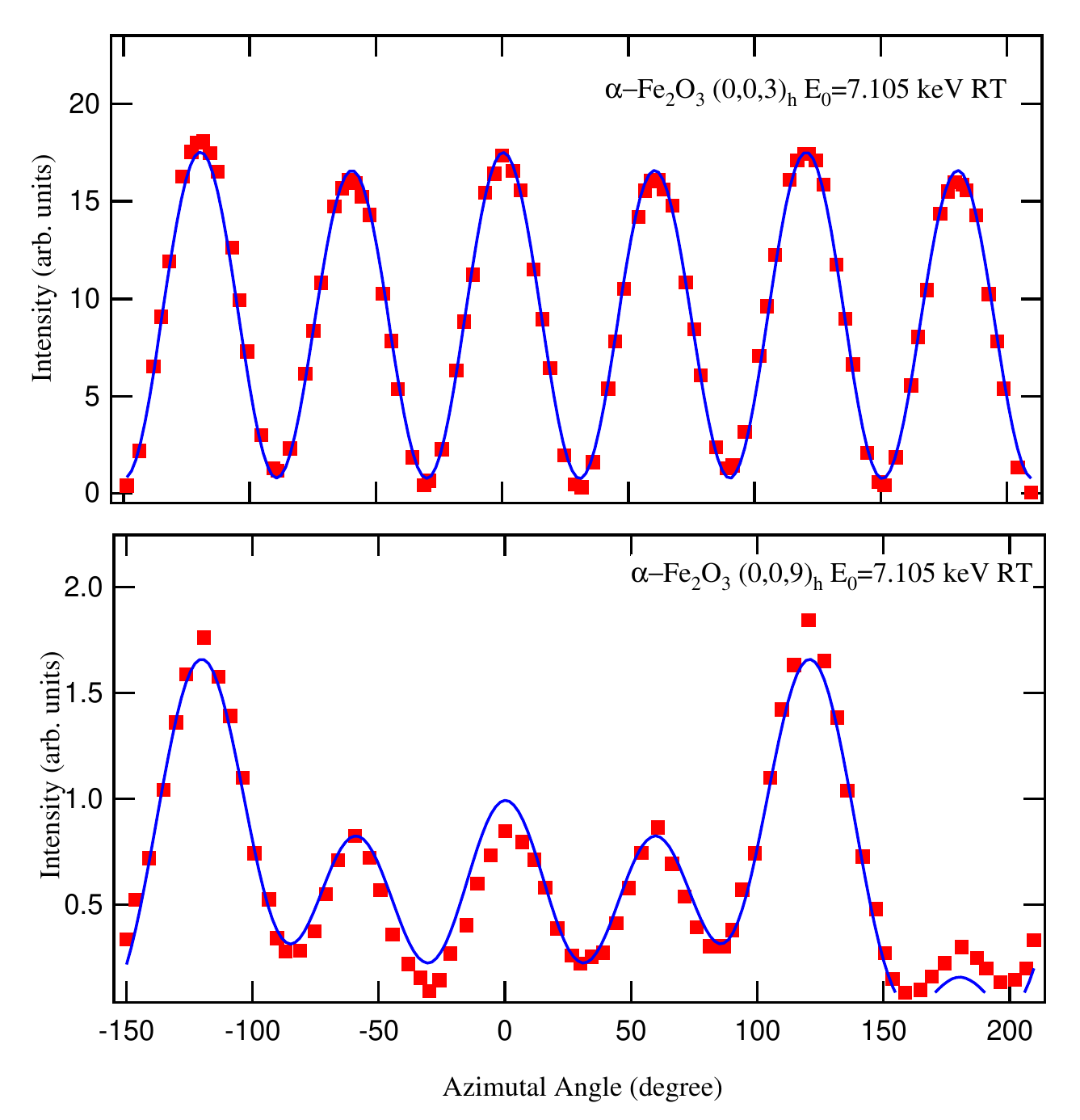}
\caption{\label{fig:figure4} Azimuthal-angle dependence of
intensity of Bragg reflections $(0,0,l)_h$ with $l = 3$ and $l =
9$ for phase II (room temperature). Continuous curves are fits to
structure factors for E1-E2 and E2-E2 events with magnetic
(time-odd) parity-even multipoles set to zero. Inferred relative
atomic multipoles are listed in Table \ref{tab:table1}.
Experimental data is taken from Kokubun $\it et\, al
$.\cite{Kokubun1}}
\end{center}
\end{figure}

Figure \ref{fig:figure4} shows fits of E1-E2 and E2-E2 structure
factors to data gathered at $l = 3$ and $l = 9$ in phase II (room
temperature). As before, in our analysis of data gathered on phase
I, parity-even multipoles with odd K are set to zero. Time-even
contributions to structure factors, determined by chemical
structure, are taken to be the same in phases I and II.
Consistency with this assumption, about chemical structure,
implies for phases I and II the same values of $\langle
T^4_{+3}\rangle'$ and $\langle U^2_0\rangle$. Inferred relative
values of time-odd atomic multipoles for phase II are listed in
Table \ref{tab:table1}, with values of $\langle T^4_{+3}\rangle'$
and $\langle U^2_0\rangle$ in the ratio $20: 1$. Relative to the
magnitude of $\langle U^2_0\rangle$, none of the magneto-electric
multipoles are negligible in phase II. Figure \ref{fig:figure3}
contains a fit of pure parity-even structure factors, E1-E1 and
E2-E2, to data for the reflection l = 9, and the quality of the
fit is clearly inferior to that reported in figure
\ref{fig:figure4} with E1-E2 and E2-E2 structure factors.

\section {Chiral state}

A chiral, or handed, state of a material is permitted to couple to
a probe with a like property, in our case circular polarization
(helicity) in the primary beam of x-rays. In our notation, the
pseudo-scalar for helicity, $P_2$, is one of three purely real,
time-even Stokes parameters. Intensity induced by helicity in the
primary beam is,\cite{Fernandez17}

\begin{equation}
 I_c = P_2 \,Im \,\{G^*_{\sigma'\pi}\, G_{\sigma'\sigma} + G^*_{\pi'\pi}\, G_{\pi'\sigma}\},
\label{Eq5}
\end{equation}

where the amplitudes $G_{\mu'\nu}$ are given by eq. (\ref{Eq1})
and
* denotes complex conjugation.  $I_c$ is zero for Thomson scattering
since it is proportional to $({\bf e} \cdot {\bf e'})$ and
diagonal with respect to states of polarization.

Let us consider the fully compensating collinear antiferromagnet
(phase I). For both E1-E1 and E1-M1 events there are no
contributions diagonal with respect to states of polarization and
$I_c$ is zero. Using structure factors listed in the Appendix for
the E1-E2 and E2-E2 events we find,

\begin{eqnarray}
 I_c(E1-E2)= -P_2\,(\frac {8 \sqrt{2}}{5})\rho^2 (E1-E2) \mid d(E)\mid ^2 sin(3 \psi)\,cos^3 (\theta)\,(1+sin^2(\theta))cos^2 (\varphi l) \langle G^3_{+3}\rangle'\langle U^2_0 \rangle,
\label{Eq6}
\end{eqnarray}

and,

\begin{eqnarray}
 I_c(E2-E2)= -P_2 \,4 \rho ^2 (E2-E2) \mid d(E) \mid ^2 \,sin(6 \psi) sin(\theta)\,cos^6 (\theta)\,sin^2 (\varphi l)\langle T^3_{+3}\rangle''  \langle T^4_{+3} \rangle',
\label{Eq7}
\end{eqnarray}

\begin{figure}[h]
\begin{center}
\includegraphics[width=10cm]{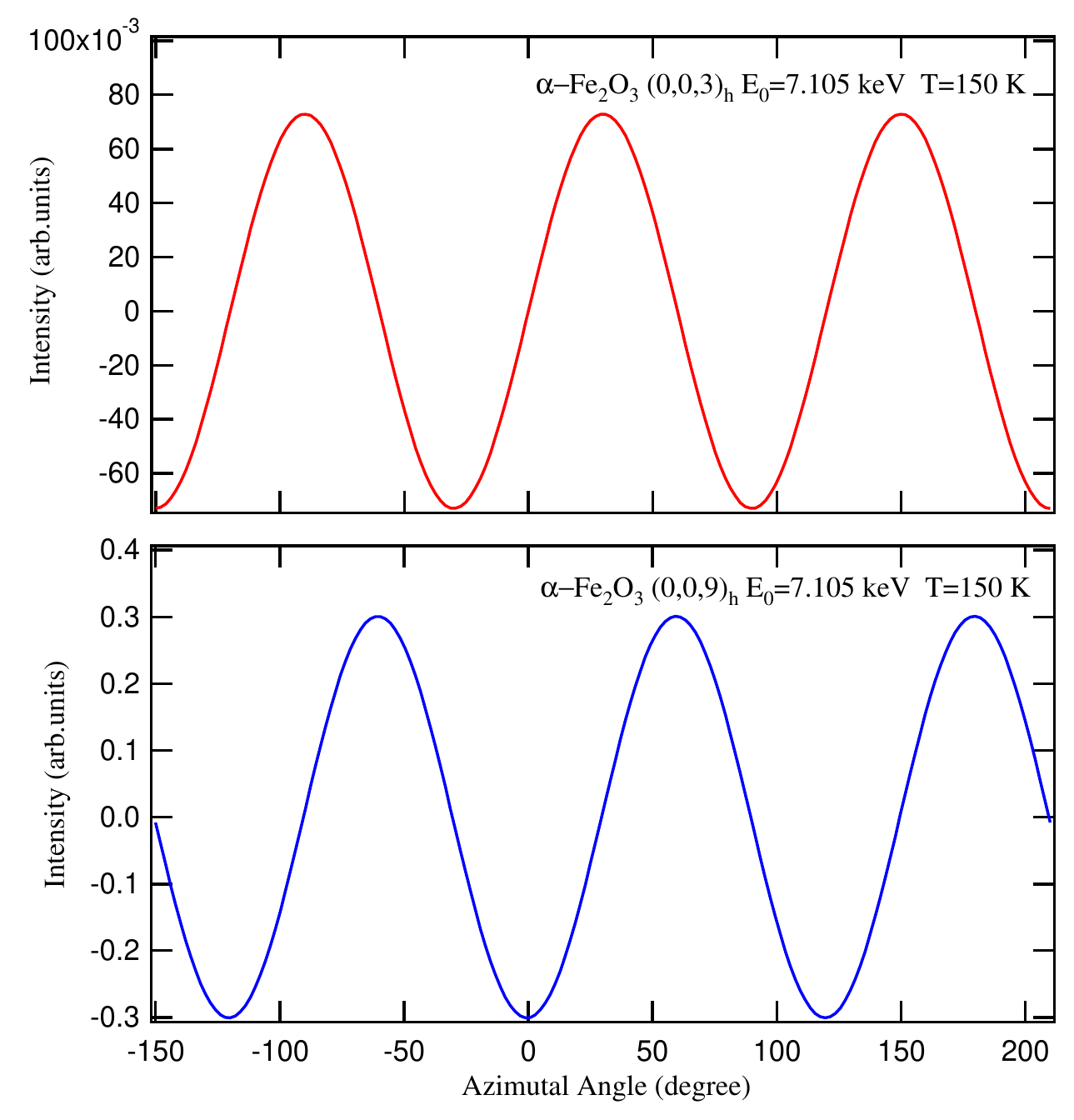}
\caption{\label{fig:figure5} Simulation of the azimuthal-angle
dependence from eq. (\ref{Eq6}) for a circular polarized light of
Bragg reflections $(0,0,l)_h$ with $l = 3$ and $l = 9$ for Phase
I. Continuous curves are simulations made with the values of the
multipoles from the E1-E2 event gathered in Table
\ref{tab:table1}. For the E2-E2 event $I_c$ is zero because our
magnetic (time-odd) parity-even multipoles are zero for a ferric
ion. Zero $I_c$ does not mean zero intensity for Ic is only the
circular polarization contribution to
intensity.\cite{Fernandez17}}
\end{center}
\end{figure}

The predicted intensities are significantly different - notably in
dependence on the azimuthal angle - and offer a method by which to
distinguish contributions from the two events. Intensities
(\ref{Eq6}) and (\ref{Eq7}) depend on long-range magnetic order,
with $I_c(E2-E2) = 0$ if the ferric ion is pure $^6$S. The polar
quadrupole in (\ref{Eq6}) is a manifestation of local chirality
\cite{Lovesey19, Dmitrienko25} whereas the pseudoscalar $\langle
U^0_0\rangle$ , discussed in the next section, is a conventional
measure of the chirality of a material. While for phase II, we
find that Ic is given by,

\begin{widetext}
\begin{eqnarray}
I_c(E1-E2)= P_2 \,(\frac {8 \sqrt{2}}{5})\,\rho^2 (E1-E2)\,\mid
d(E)\mid ^2 \,cos^2(\varphi l)\,cos^2(\theta )\,\langle U^2_0
\rangle \{\frac{1}{\sqrt{3}} \,sin(\psi)
\,[\frac{-3}{\sqrt{5}}\,(cos(3 \theta)+cos(\theta))\,\langle
G^1_{+1} \rangle' +\nonumber\\ +  (cos(3 \theta)-cos(\theta)) \,
\langle G^2_{+1}\rangle''- \frac{1}{\sqrt{5}}\,
(cos^3(\theta)+2cos(\theta))\, \langle G^3_{+1}\rangle']
-sin(3\psi)\,cos(\theta)\,(1+sin^2(\theta)) \langle G^3_{+3}\rangle'\},\nonumber\\
 \label{Eq8}
\end{eqnarray}

\begin{eqnarray}
I_c(E2-E2)=-P_2(\frac {1} {\sqrt{2}}) \,  \rho^2 (E2-E2)\, \mid
d(E)\mid ^2 \,sin^2(\varphi l)\, \langle T^4_{+3}\rangle'
 \{ 4\,sin(\psi)\,cos^4(\theta)\,[\frac{-1}{\sqrt{5}}\,
sin(\theta)\,(8cos^2(\theta)-5) \langle T_1^1 \rangle''
+\nonumber\\ +\sqrt{\frac{3}{5}}\, sin(\theta )cos^3 (\theta)
\langle
T^3_{+1}\rangle'']-4\sqrt{2}\,sin(\theta)\,cos^6(\theta)\,sin(6\psi)\,\langle
T^3_{+3} \rangle''\},\nonumber\\ \label{Eq9}
\end{eqnarray}
\end{widetext}

\begin{figure}[h]
\begin{center}
\includegraphics[width=10cm]{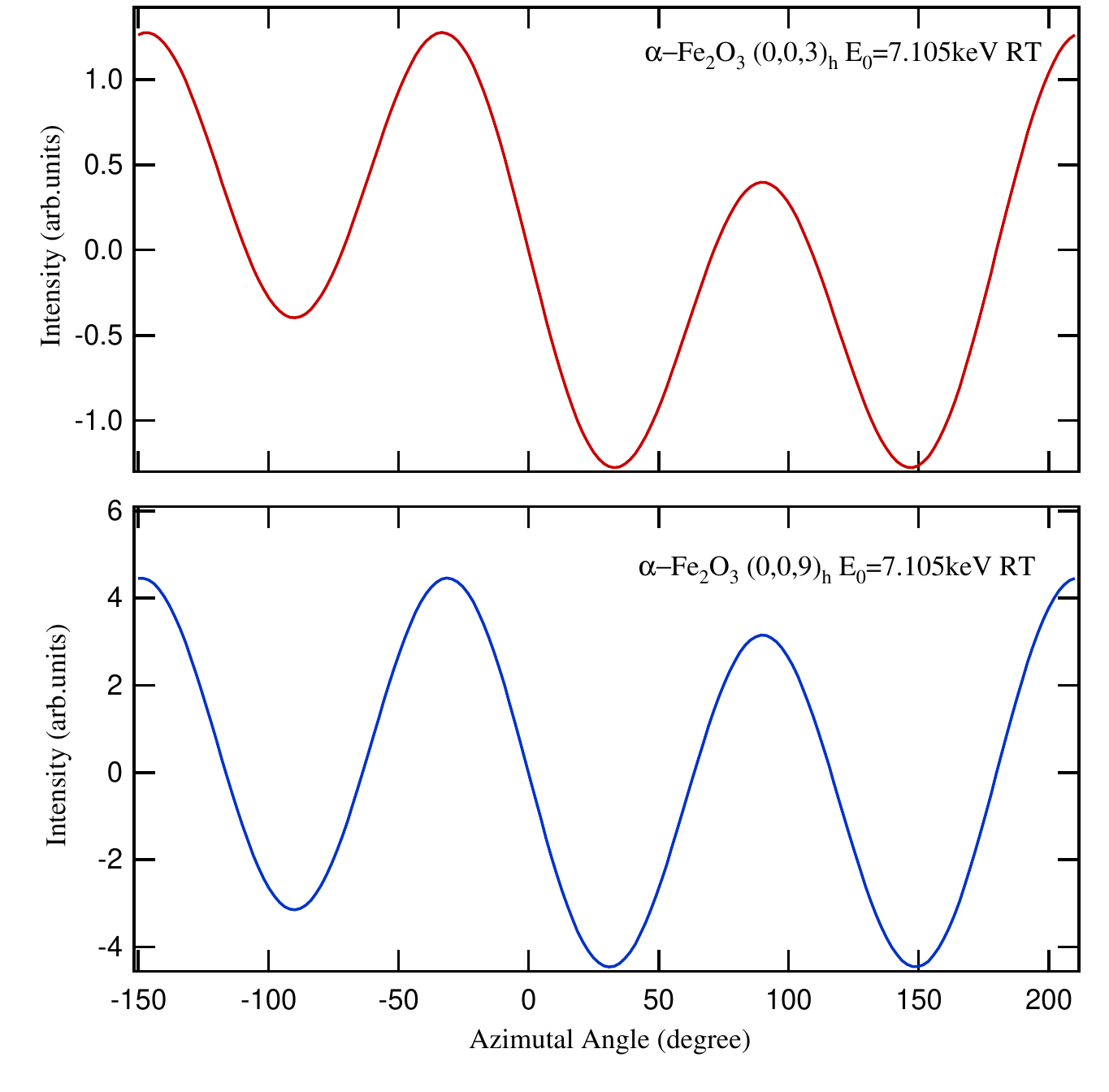}
\caption{\label{fig:SemFD-FH} Simulation of the azimuthal-angle
dependence from eq. (\ref{Eq8}) for a circular polarized light of
Bragg reflections $(0,0,l)_h$ with $l = 3$ and $l = 9$ for Phase
II (room temperature). Continuous curves are simulations made with
the values of the multipoles from the E1-E2 event gathered in
Table \ref{tab:table1}. For the E2-E2 event the Ic is equal to
zero because our magnetic (time-odd) parity-even are zero. Zero
$I_c$ does not mean zero intensity since $I_c$ is only the
circular polarization contribution.\cite{Fernandez17}}
\end{center}
\end{figure}

\section {Magnetic charge and chirality}
The pseudo-scalar monopoles  $\langle G^0_0\rangle $  and $\langle
U^0_0\rangle $ have particularly simple and interesting physical
interpretations. Both monopoles are allowed in haematite structure
factors for the E1-M1 event, as we see by inspection of relevant
expressions in the Appendix. A conventional measure of the
chirality of electrons in a molecule or extended media is $
\langle {\bf S} \cdot {\bf p}\rangle $ / $ \mid \langle {\bf
p}\rangle \mid $, where $ {\bf S} $ and $ {\bf p} $ are operators
for spin and linear momentum, and, not unsurprisingly, $\langle
U^0_0\rangle $ is proportional to  $ \langle {\bf S} \cdot {\bf p}
\rangle /\mid {\bf p} \mid $. It is well-known that, $\langle
U^0_0\rangle $ contributes to natural circular
dichroism.\cite{Collins26} On the other hand, $\langle
G^0_0\rangle$ , a magnetic charge, does not contribute to dichroic
signals but it can contribute in scattering. Such is the case for
gallium ferrate,\cite{Staub27} and phase I of haematite. Magnetic
charge, and the magneto-electric quadrupole, are present in the
amplitude for back-scattering with $ {\bf q} = - {\bf q'}$.

\section {Discussion}

We report successful analyses of resonant Bragg diffraction data
gathered by Kokubun $\it et \, al.$ \cite{Kokubun1} on haematite
in the collinear (phase I) and canted (phase II) antiferromagnetic
phases, with no analysis of diffraction according to polarization
of the x-rays. We infer good estimates of iron atomic multipoles,
and find large amounts of parity-odd multipoles. Of particular
importance to a successful analysis is a polar quadrupole, a
measure of local chirality,\cite{Dmitrienko25} and, in phase II,
magneto-electric multipoles that include the anapole. Slight
departures between our theory and experiment could be due to a
less than ideal crystal, as witnessed in the extended interval of
temperature for rotation of magnetic moments between phases I and
II.\cite{Kokubun1}

Future experiments might employ polarization analysis that will
allow closer scrutiny of unit-cell structure factors for haematite
we list in an Appendix, which are derived from the established
chemical and magnetic structures of haematite. We predict for
phase I that scattering enhanced by the E1-M1 event contains
monopoles that represent chirality and magnetic charge.

Our analyses of data are based on an atomic theory of x-ray Bragg
diffraction\cite{Lovesey19} with unit-cell structure factors that
are fundamentally different from corresponding structure factors
employed by Kokubun $\it et \, al.$\cite{Kokubun1} One difference
arises in the treatment of non-resonant magnetic scattering. We
use the exact expression, due solely to spin moments, while
Kokubun $\it et \, al.$\cite{Kokubun1} mistakenly - because it is
not valid in the investigated interval of energy - use an abridged
amplitude by de Bergevin and Brunel\cite{Bergevin22} that is a sum
of the exact expression and the high-energy limit of two
contributions to scattering that involve intermediate states (one
of the two is capable of showing a resonance).Treating the
resonance as a single oscillator, in accord with the reported
energy profile, our structure factors for resonant diffraction are
completely determined with no arbitrary phase factors, unlike the
analysis in Ref. (\onlinecite{Kokubun1}). This difference in the
analyses is a likely explanation of our evidence that published
data for azimuthal-angle scans are miss-set by $30$ deg. Our
treatment of magnetic (time-odd) contributions to scattering is
another major difference in the analyses. Whereas Kokubun $\it et
\, al.$\cite{Kokubun1} allow only the dipole in the E1-E1 event we
consider all permitted time-odd contributions in both parity-even
and parity-odd events. Time-odd multipoles from parity-even
events,  $ \langle T^K_Q\rangle $ with odd K, are related to
orbital magnetism when the intermediate state in resonance is an
s-state, as is the case in the experiments in question with
absorption at the iron K-edge. The available evidence is that
orbital magnetism of the ferric ion in haematite is negligible, as
expected for an s-state ion, and the same can be said of the
parity-even, time-odd multipoles, including the dipole which at
resonance is the only source of magnetic scattering considered in
Ref. (\onlinecite{Kokubun1}). From our analysis, we conclude that
magnetic scattering at resonance is provided by magneto-electric
multipoles in an E1-E2 event. We demonstrate beyond reasonable
doubt that, allowing magnetic  $ \langle T^K_Q\rangle $ different
from zero the available data are not consistent with diffraction
enhanced by purely parity-even events, E1-E1 and E2-E2.

In summary, we have derived information on the relative magnitude
of multipoles for the antiferromagnetic phases of haematite (above
and below the Morin Temperature) These estimates are obtained from
analyses of experimental azimuthal dependence gathered in resonant
x-ray Bragg diffraction at space-group forbidden reflections
$(0,0,3)_h $ and $(0,0,9)_h $. A chiral electron state is proposed
from a predicted coupling of resonant intensity to circular
polarization in the primary beam. This effect allows
differentiating between contributions of the E1-E2 and E2-E2
events. In addition, pseudo-scalar monopoles (chirality and
magnetic charge) are present in the E1-M1 amplitude for resonant
scattering by haematite below the Morin temperature.

\section{ACKNOWLEDGMENTS}
Professor Gerrit van der Laan provided values of atomic radial
integrals for a ferric ion. We have benefited from discussions
with Dr. A. Bombardi and Professor S. P. Collins, and
correspondence with Dr. F. de Bergevin. One of us (SWL) is
grateful to Professor E. Balcar for ongoing noetic support.
Financial support has been received from Spanish FEDER-MiCiNN
Grant No. Mat2008-06542-C04-03. One of us A.R.F is grateful to
Gobierno del Principado de Asturias for the financial support from
Plan de Ciencia, Tecnología e innovación (PTCI) de Asturias.

\appendix
\section{ Unit-cell structure factors}
Some factors in eq.(\ref{Eq1}) contain a dimensionless quantity
$\aleph = m\Delta a_o^2/\hbar^2 = 260.93$ where $a_o$ is the Bohr
radius and $\Delta = 7.105 keV$. Radial integrals for the E1 and
E2 processes at the K-absorption edge are denoted by $\{R\}_{sp}$
and $\{R^2\}_{sd}$. Estimates from an atomic code are
$\{R\}_{1s4p}/a_o = -0.0035$ and $ \{ R^2 \}_{1s3d}/a_o^2 =
0.00095 $, and it is interesting that the magnitudes are smaller
than hydrogenic values with $Z = 26$ by a factor of about three.
More appropriate values of the radial integrals will be influenced
by ligand ions. The M1 process between stationary states of an
isolated non-relativistic ion is forbidden because the radial
overlap of initial and final states in the process is zero, on
account of orthogonality. For an M1 process in a compound the
radial integral, denoted here by $\{1\}_{\gamma\gamma}$ , is an
overlap of two orbitals with common orbital angular momentum,
$\Gamma$, which may be centred on different ions. The magnitude of
$\{1\}_{\gamma\gamma}$ is essentially a measure of configuration
interactions and bonding, or covalancy, of a cation and ligands.
Factors appearing in eq.(\ref{Eq1}) are,

\begin{eqnarray}
 \rho(E1-E1) = [ \{R \}_{sp}/a_o]^2\aleph ,
\end{eqnarray}

\begin{eqnarray}
 \rho(E1-M1) = q \{R\}_{sp}\{1\}_{\gamma\gamma}  ,
\end{eqnarray}

\begin{eqnarray}
 \rho(E1-E2) = [q \{R^2\}_{sd}{R}_{sp}/a_o^2]\aleph ,
\end{eqnarray}

\begin{eqnarray}
 \rho(E2-E2) = [q \{R^2 \}_{sd}/a_o]^2\aleph .
\end{eqnarray}

 Haematite structure factors $F_{\mu'\nu} $ for
forbidden reflections $(0,0,l)_h$ with $l = 3(2n + 1)$ and
enhancements by E1-E1, E1-M1, E1-E2 and E2-E2 events are listed
below. In these expressions, the angle $\varphi = - \pi u $, where
$u=2z-1/2=0.2104$ for $\alpha-Fe_2O_3$, the angle $\theta$ is the
Bragg angle, and $\langle T_Q^K \rangle $, $\langle G_Q^K \rangle
$ and $\langle U_Q^K \rangle$ are the mean values of the atomic
tensors involved.

\subsection{\label{app:subsec} Collinear antiferromagnet, phase I}

(E1-E1)
\begin{eqnarray}
 F_{\sigma'\sigma}(E1-E1)=0 \label{appa}
\end{eqnarray}
\begin{eqnarray}
F_{\pi'\sigma}(E1-E1)=-2\sqrt{2}sin(\varphi l)sin(\theta) \langle
T^1_0\rangle \label{appb}
\end{eqnarray}
\begin{eqnarray}
F_{\pi'\pi}(E1-E1)=0
 \label{appc}
\end{eqnarray}

(E1-M1)

\begin{eqnarray}
F_{\sigma'\sigma}(E1-M1)=0 \label{appd}
\end{eqnarray}
\begin{eqnarray}
F_{\pi'\sigma}(E1-M1) = \frac {2\sqrt{2}} {\sqrt{3}} cos(\varphi
l) \{2\sqrt{2}[-sin^2(\theta) \langle G^0_0 \rangle +i
cos^2(\theta) \langle U^0_0 \rangle] +(2+cos^2(\theta)) \langle
G^2_0 \rangle + i cos^2(\theta) \langle U^2_0 \rangle\}
\label{appe}
\end{eqnarray}
\begin{eqnarray}
F_{\pi'\pi}(E1-M1)=0 \label{appf}
\end{eqnarray}

(E1-E2)

\begin{eqnarray}
F_{\sigma'\sigma}(E1-E2)= -\frac {4\sqrt{2}} {\sqrt{5}}sin(3
\psi)cos(\varphi l) cos (\theta)\langle G^3_{+3} \rangle'
\label{appg}
\end{eqnarray}
\begin{eqnarray}
F_{\pi'\sigma}(E1-E2)= \frac {2}{\sqrt{5}} cos(\varphi l)
\{-[3cos^2(\theta)-2] \langle G^2_0\rangle + i cos^2(\theta)
\langle U^2_0 \rangle
 +\sqrt{2} sin(2\theta) cos(3\psi) \langle G^3_{+3} \rangle'\}
 \label{apph}
\end{eqnarray}
\begin{eqnarray}
F_{\pi'\pi}(E1-E2)= -\frac {4\sqrt{2}} {\sqrt{5}} sin(3
\psi)cos(\varphi l)  cos(\theta)sin^2(\theta)\langle G^3_{+3}
\rangle'
 \label{appi}
\end{eqnarray}

(E2-E2)

\begin{eqnarray}
F_{\sigma'\sigma}(E2-E2)= -\sqrt{2} sin(3 \psi) sin(\varphi
l)\langle T^3_{+3}\rangle'' \label{appj}
\end{eqnarray}
\begin{eqnarray}
F_{\pi'\sigma}(E2-E2)=  \sqrt{\frac{2}{5}}sin(\varphi l) \{
sin(3\theta) \langle T^1_0 \rangle  -
sin(\theta)[3cos^2(\theta)-2] \langle T^3_0 \rangle -\nonumber\\
-\frac {\sqrt{5}} {4} cos(3\psi) [[3cos(3\theta)+cos(\theta)]
\langle T^3_{+3}\rangle''-i[cos(3\theta)+3cos(\theta)] \langle
T^4_{+3} \rangle']\}
 \label{appk}
\end{eqnarray}
\begin{eqnarray}
 F_{\pi'\pi}(E2-E2)= -\frac {1} {\sqrt{2}} sin(3\psi) sin(\varphi l)  sin(4\theta)\langle T^3_{+3} \rangle''
  \label{appl}
\end{eqnarray}

\subsection{\label{app:subsec}Canted antiferromagnet, phase II}

Time-even contributions to structure factors, determined by
chemical structure, are the same in phases I and II. Thus the
structure factor with polar multipoles, $F_{\mu'\nu}(u)$, for
phase II is identical to the foregoing expression for phase I.
With the convenience of the reader in mind, structure factors for
parity-even multipoles,  $F_{\mu'\nu}(t)$, are given in full
although contributions only with K = 1 and 3 differ from foregoing
expressions.
\\

(E1-E1)
\begin{eqnarray}
F_{\sigma'\sigma}(E1-E1)=0 \label{bppa}
\end{eqnarray}
\begin{eqnarray}
F_{\pi'\sigma}(E1-E1) = 4 cos(\psi)sin(\varphi l)
cos(\theta)\langle T^1_{+1} \rangle'' \label{bppb}
\end{eqnarray}
\begin{eqnarray}
F_{\pi'\pi}(E1-E1)= 4 sin(\psi) sin(\varphi l)
sin(2\theta)\langle T^1_{+1} \rangle'' \label{bppc}
 \end{eqnarray}

(E1-M1)
\begin{eqnarray}
F_{\sigma'\sigma}(E1-M1)=8 sin(\psi)cos(\varphi l)
cos(\theta[-\langle G^1_{+1} \rangle'+\langle
G^2_{+1}\rangle'']\label{bppd}
\end{eqnarray}
\begin{eqnarray}
F_{\pi'\sigma}(E1-M1)= 4 cos(\psi)cos(\varphi l) sin (2\theta) [
\langle G^1_{+1} \rangle'] \label{bppe}
\end{eqnarray}
\begin{eqnarray}
F_{\pi'\pi}(E1-M1)= -8 sin(\psi)cos(\varphi l) cos(\theta)[\langle
G^1_{+1}\rangle'+\langle G^2_{+1} \rangle''] \label{bppf}
\end{eqnarray}

(E1-E2)
\begin{eqnarray}
F_{\sigma'\sigma}(E1-E2)= \frac{4\sqrt{2}} {\sqrt{5}} cos(\varphi
l) cos (\theta)\ \{ \frac {1} {\sqrt{3}} sin(\psi) [\frac {-3}
{\sqrt{5}} \langle G^1_{+1} \rangle'-\langle G^2_{+1}\rangle''
+\frac {1} {\sqrt{5}} \langle G^3_{+1}\rangle']- sin(3\psi)
\langle G^3_{+3} \rangle'\} \label{bppg}
\end{eqnarray}
\begin{eqnarray}
F_{\pi'\sigma}(E1-E2)= 2\sqrt{\frac {2} {5}}  cos(\varphi
l)sin(2\theta) \{\frac {cos(\psi)} {\sqrt{3}} [\frac {3}
{\sqrt{5}} \langle G^1_{+1} \rangle'-2\langle G^2_{+1} \rangle'' -
\frac {1}{\sqrt{5}} \langle G^3_{+1} \rangle']-cos(3\psi)\langle
G^3_{+3} \rangle'\} \label{bpph}
\end{eqnarray}
\begin{eqnarray}
F_{\pi'\pi}(E1-E2)= -\frac {4\sqrt{6}} {5}  cos(\varphi l)\{
\sqrt{\frac{5} {3}} cos(\theta)sin^2(\theta)sin(3\psi) \langle
G^3_{+3} \rangle' +\nonumber\\
+ sin(\psi) [cos(3\theta) [\langle G^1_{+1} \rangle' - \frac
{\sqrt{5}}{3} \langle G^2_{+1} \rangle'' ] +\frac {1} {3}
[cos^3(\theta)+3 cos(\theta) ]\langle G^3_{+1} \rangle'] \}
\label{bppi}
\end{eqnarray}

(E2-E2)
\begin{eqnarray}
 F_{\sigma'\sigma}(E2-E2)= sin(2\theta)  sin(\varphi l) \{sin(\psi)[\frac {-2} {\sqrt{5}} \langle T^1_{+1}\rangle''-\sqrt{\frac {6} {5}} \langle T^3_{+1} \rangle'']+ \sqrt{2}sin(3\psi)\langle T^3_{+3} \rangle''\} \label{bppj}
\end{eqnarray}
\begin{eqnarray}
F_{\pi'\sigma}(E2-E2)= - sin(\varphi l)\{ cos(\psi) \times[\frac
{2}{\sqrt{5}}  cos(3\theta) \langle
T^1_{+1}\rangle''+\sqrt{\frac{6} {5}} cos(\theta) (1+sin^2(\theta)
)\langle T^3_{+1} \rangle''] +\nonumber\\ +cos(3\psi)[\sqrt{2}
cos(\theta) (3 cos^2(\theta)-2) \langle T^3_{+3} \rangle'' ]\}
\label{bppk}
\end{eqnarray}
\begin{eqnarray}
F_{\pi'\pi}(E2-E2)= \frac{1}{\sqrt{2}}  sin(\varphi l)
sin(4\theta) \{sin(\psi) [ - \frac {4 \sqrt{2}}{\sqrt{5}} \langle
T^1_{+1} \rangle''+ \sqrt{\frac{3}{5}} \langle T^3_{+1} \rangle''
] -sin(3\psi) \langle T^3_{+3} \rangle''\} \label{bppl}
\end{eqnarray}

\bibliography{Fe2O3bib}

\end{document}